\documentclass[11pt]{article}

\usepackage[margin=1in]{geometry}
\usepackage[T1]{fontenc}
\usepackage[utf8]{inputenc}
\usepackage{lmodern}
\usepackage{microtype}
\usepackage{setspace}
\usepackage{booktabs}
\usepackage{tabularx}
\usepackage{ragged2e}
\usepackage{amsmath,amssymb}

\usepackage{graphicx}
\usepackage{caption}
\usepackage{siunitx}
\usepackage[version=4]{mhchem}

\usepackage[hidelinks]{hyperref}

\usepackage{titlesec}
\titleformat{\section}{\large\bfseries}{\thesection.}{0.6em}{}
\titleformat{\subsection}{\normalsize\bfseries}{\thesubsection.}{0.6em}{}
\titlespacing*{\section}{0pt}{10pt}{6pt}
\titlespacing*{\subsection}{0pt}{8pt}{4pt}

\usepackage{authblk}

\title{\bfseries Automated multiphase identification and refinement in powder diffraction using mismatch-tolerant machine learning}

\author[1]{Lalit Yadav\thanks{Corresponding author: \texttt{yadavl@ornl.gov}}}
\author[1]{Yongqiang Cheng}
\author[1]{Mathieu Doucet}

\affil[1]{Neutron Scattering Division, Oak Ridge National Laboratory, Oak Ridge, TN 37831, USA}

\date{}

\usepackage[numbers,sort&compress]{natbib}

\usepackage[switch]{lineno}

\begin{document}

\begin{titlepage}
\thispagestyle{empty}
\begin{center}
\vspace*{\fill}
\begin{minipage}{0.9\textwidth}
\small
This manuscript has been authored in part by UT-Battelle, LLC, under contract
DE-AC05-00OR22725 with the U.S. Department of Energy (DOE). The U.S. Government
retains, and the publisher, by accepting the article for publication, acknowledges
that the U.S. Government retains a nonexclusive, paid-up, irrevocable, worldwide
license to publish or reproduce the published form of this manuscript, or allow
others to do so, for U.S. Government purposes. DOE will provide public access to
these results of federally sponsored research in accordance with the DOE Public
Access Plan (\url{https://energy.gov/downloads/doe-public-access-plan}).
\end{minipage}
\vspace*{\fill}
\end{center}
\end{titlepage}

\maketitle

\begin{abstract}
Powder diffraction is a primary structural characterization tool in materials science, yet automated phase identification remains a major bottleneck for autonomous discovery. Existing workflows rely heavily on search--match heuristics and manual Rietveld refinement, and broadly usable end-to-end automation is especially limited for neutron powder diffraction, where comparable tools are largely absent. Here we introduce RADAR-PD, a modality-aware machine learning framework for phase identification and quantification across both X-ray and neutron powder diffraction. RADAR-PD couples a mismatch-tolerant neural network operating on coarse momentum-transfer fingerprints with automated lattice nudging and physics-constrained Rietveld verification, enabling dominant-phase hypotheses to be generated from elemental constraints and secondary phases to be isolated recursively. On an experimental RRUFF PXRD benchmark, RADAR-PD outperforms DARA in recovering the reference phase. RADAR-PD further provides robust multiphase analysis on complex time-of-flight and constant-wavelength neutron datasets, addressing an important unmet need in automated neutron diffraction analysis. These results establish RADAR-PD as an auditable, instrument-agnostic framework for autonomous structural discovery.
\end{abstract}


\section{Introduction}

Powder diffraction is a fundamental technique for phase analysis across neutron and X-ray facilities, enabling quantitative studies of structure--property relationships, phase transformations, and ordering phenomena. Increasingly, diffraction experiments are moving toward streaming analysis and closed-loop experiment steering, where structural-property-based decisions must be made while data are actively being acquired.\cite{Allan2019BlueskyAhead,McDannald2022ANDiE,ThayerILLUMINE_LCLStream,LCLS_ILLUMINE,OLCF2024_StreamingPipeline}. In this setting, full-profile Rietveld refinement\cite{Rietveld1969,McCusker1999RietveldGuidelines} provides a physically grounded model for interpreting diffraction histograms and extracting quantitative phase fractions \cite{HillHoward1987NeutronQPA}. Modern open-source refinement toolchains, such as GSAS-II \cite{TobyVonDreele2013GSASII,ODonnell2018GSASIIAPI}, further enable the scripting and high-throughput operation necessary for facility-scale workflows.

A persistent bottleneck arises when a measured diffraction pattern contains reflections not explained by the anticipated primary-phase model. Such extra peaks can indicate genuine physics (e.g., symmetry lowering or additional ordering) but can also reflect secondary phases introduced during synthesis, handling, or sample environment. These unidentified features therefore create a critical ambiguity: they obscure whether the signal arises from novel intrinsic phenomena or extrinsic contamination, while simultaneously confounding quantitative interpretation of the primary phase. In routine practice, phase identification often begins with commercial peak-list search--match against proprietary reference libraries (e.g., PDF-5+) \cite{Kabekkodu2024PDF5plus,CrystalImpactMatch,Degen2014HighScoreSuite}. However, these workflows are PXRD-centric and do not readily extend to neutron or synchrotron/beamline X-ray data; in practice, analysts often fall back on intuition-guided manual database search \cite{Hellenbrandt2004ICSD,Zagorac2019ICSDDevelopments,Grazulis2012COD,Jain2013MaterialsProject} and repeated trial multiphase refinements. This workflow is time-consuming, difficult to reproduce, and brittle when impurity peaks are weak, overlap dominant reflections, or shift due to lattice mismatch between database entries and experimental conditions. Crucially, this manual intervention fractures end-to-end automation, limiting both ubiquitous laboratory X-ray diffractometers and central neutron/X-ray user facilities where on-the-fly decision-making is essential.

From a computational perspective, rapid phase discovery is a massive pattern-matching problem under distribution shift. Experimental patterns depend on instrument resolution, radiation modality, background and sample environment, counting statistics, and peak overlap; candidate structures come from databases with lattice parameters and settings that may not match experimental conditions. Several approaches address parts of this problem, but practical gaps remain. Full-profile search--match methods based on automated Rietveld refinement avoid peak picking and can support identification and quantification, but require repeated refinements over many candidates and can become costly as the admissible chemical space or reference set grows \cite{Lutterotti2019FPSM}. Automation frameworks such as XERUS \cite{BaptistaDeCastro2022XERUS} and Dara \cite{Fei2026Dara} accelerate parts of the process by combining candidate screening with refinement-based validation. However, because these strategies ultimately depend on simulating and comparing candidate profiles across large reference sets (and, for multi-phase problems, across combinations), runtime scales with the number of candidate references and the size of the hypothesis space. In addition, severe peak overlap and database--experiment lattice mismatch remain challenging in practice, destabilizing screening and downstream refinement \cite{BaptistaDeCastro2022XERUS,Fei2026Dara}.

Conversely, machine-learning (ML) approaches provide major speedups but are often confined to restricted, closed-set settings. Deep-learning classifiers and retrieval systems such as PQ-Net \cite{Dong2021PQNet}, CPICANN \cite{Zhang2024CPICANN}, and XQueryer\cite{Cao2025XQueryer} can perform strongly within fixed catalogs and training regimes, but performance can degrade under complex multiphase mixtures and background variability, and adaptation to new instruments, radiation modalities, or updated reference libraries typically requires substantial retraining or regeneration of training data \cite{Lee2020SyntheticXRD_DLPhaseID,Schuetzke2021IucrjTrainingRealism}. These limitations motivate a complementary strategy: a fast, mismatch-tolerant proposer coupled with a physics-constrained verifier, enabling robust hypothesis generation that can operate across instruments and evolving databases without instrument-specific retraining, while still producing refinement-grade, auditable conclusions.

Here, we introduce RADAR-PD (Residual-Aware Deep-learning--Assisted Refinement for Powder Diffraction), a modality-aware propose--verify workflow designed for universal phase discovery. Unlike approaches that rely on exhaustive exact-pattern simulation or fixed training catalogs, RADAR-PD separates fast hypothesis generation from rigorous Reitvield refinement based verification. At its core is a compact neural scorer that operates on coarse $Q$-intensity fingerprints rather than exact diffraction profiles. This representation intrinsically tolerates modest peak shifts, multiphase overlap, and differences in instrumental resolution, enabling rapid database-scale screening without instrument-specific retraining. To address a critical failure mode in automated refinement---database-to-experiment peak-position mismatch---RADAR-PD introduces an automated lattice-nudging step that improves candidate alignment prior to verification, stabilizing convergence during the final staged multiphase refinement in GSAS-II.

RADAR-PD supports two complementary operating pathways. For completely uncharacterized samples, a composition-only mode bootstraps a primary-phase hypothesis directly from user-provided elemental constraints, requiring no initial structural model. For targeted beamline triage, RADAR-PD reframes discovery as residual explanation: it establishes a conservative baseline refinement of a known primary phase and focuses its search on the unexplained residual signal. Because the framework selects modality-specific scattering factors and reference catalogs at runtime, both pathways operate natively across neutron and X-ray powder diffraction without altering the core propose--verify logic.

We validate this modality-general framework across both X-ray and neutron diffraction. We demonstrate robustness on complex constant-wavelength (CW) and time-of-flight (TOF) neutron datasets from HIFR and SNS beamlines at ORNL\cite{Calder2018ORNL_PowderSuite,Garlea2010HB2A,Huq2019POWGEN}, where broadly usable automated analysis tools remain limited. On the experimental RRUFF PXRD benchmark, RADAR-PD outperforms DARA in both accuracy and runtime, recovering reference phases more frequently while operating substantially faster than DARA's more exhaustive candidate-search workflow (Table~\ref{tab:benchmark_summary}). By delivering reproducible, database-scale hypothesis generation together with Rietveld refinement-verified initialization on beamline-relevant timescales, RADAR-PD provides a practical engine for autonomous diffraction workflows.


\section{Results}

\subsection{Residual-explanation workflow}

\begin{figure*}[t]
    \centering
    \includegraphics[width=\linewidth]{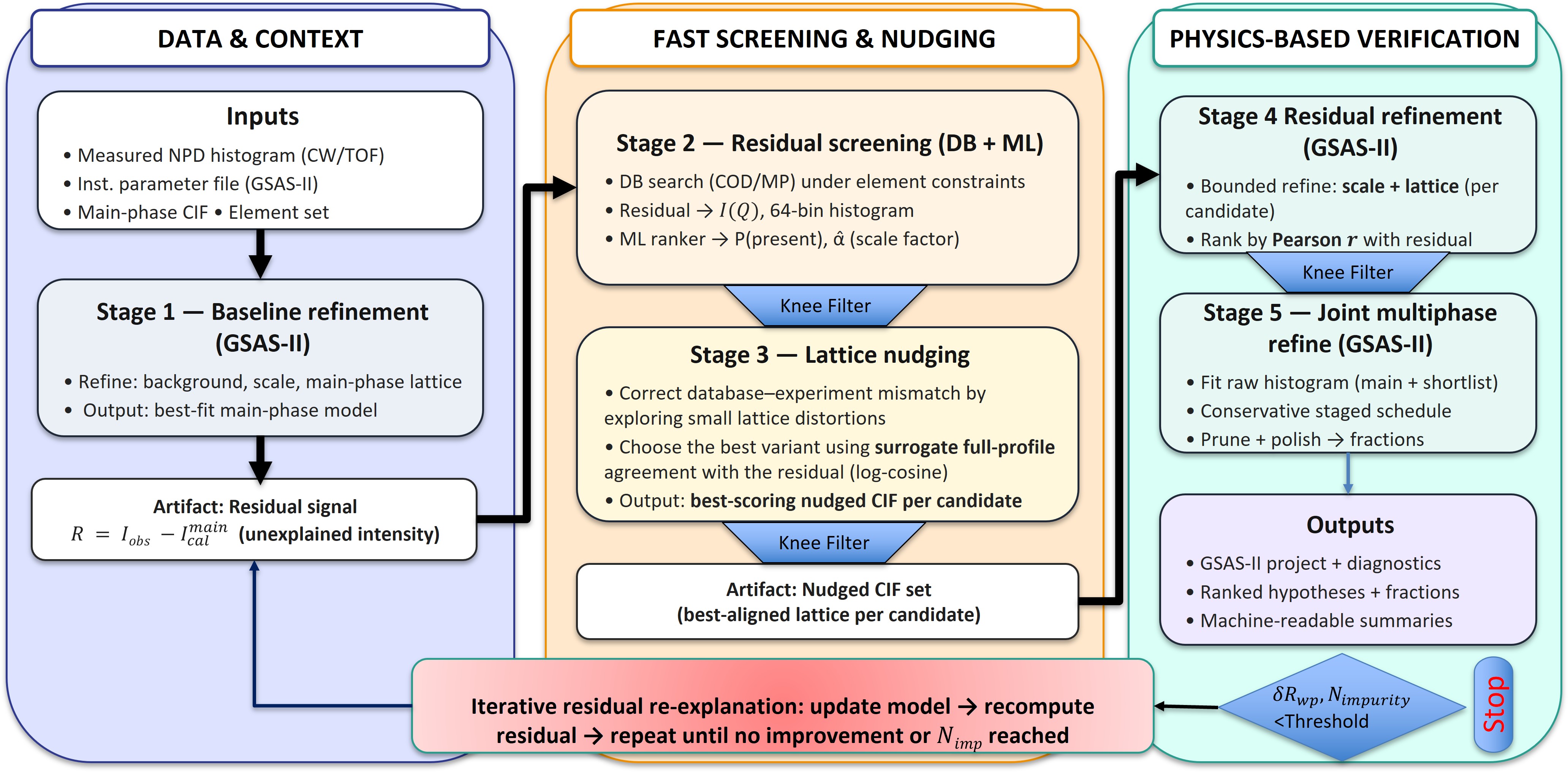}
    \caption{Workflow for the RADAR-PD. A stable baseline refinement of the main phase is used to compute a residual curve. Element-constrained database retrieval and ML ranking produce a shortlist of impurity hypotheses. Candidates are lattice-nudged to mitigate peak-position mismatch and then verified and quantified by staged multiphase refinement in GSAS-II. The loop can be repeated to explain additional residual intensity.}
    \label{fig:workflow_flowchart}
\end{figure*}

Figure~\ref{fig:workflow_flowchart} summarizes the pipeline. Starting from a measured histogram, an instrument parameter file and (in the standard mode) a main-phase CIF, we perform a conservative baseline refinement in GSAS-II. To maximize robustness, this baseline stage refines only background coefficients, the main-phase scale factor and lattice parameters. The refined main-phase model is then subtracted from the measured profile to yield a residual curve that highlights unexplained intensity.

Impurity candidates are retrieved from locally stored crystallographic databases \cite{Grazulis2012COD,Jain2013MaterialsProject} under user-provided elemental constraints. Because exhaustive simulation and refinement over all candidates is infeasible for beamline use, we screen and rank candidates using an ML residual--candidate compatibility model operating on a coarse $Q$ representation. A short list is then passed to a lattice-nudging stage to mitigate database--experiment peak-position mismatch, followed by physics-based verification in GSAS-II. Verified impurities are incorporated into the refinement model, the residual is recomputed and the search is iterated until the requested number of impurity phases is reached or the improvement saturates.

In a composition-only mode, no main-phase CIF is provided. The same screening and verification machinery is initialized from the \emph{total} measured histogram to propose a dominant phase from the allowed element set; the workflow then proceeds with the same residual-based impurity loop.

\subsection{Residual fingerprints and ML candidate ranking}

To compare experimental residuals to candidate reference patterns in an instrument-independent way, both are represented on a common momentum-transfer axis. Intensities are integrated onto a fixed grid of 64 $Q$ bins spanning $0.5 < Q < 6~\text{\AA}^{-1}$. This coarse fingerprint reduces sensitivity to modest lattice shifts, differences in instrumental broadening and imperfect subtraction of the dominant phase while preserving the distribution of scattering intensity needed for phase discrimination.

\begin{figure*}[t]
\centering
\includegraphics[width=\linewidth]{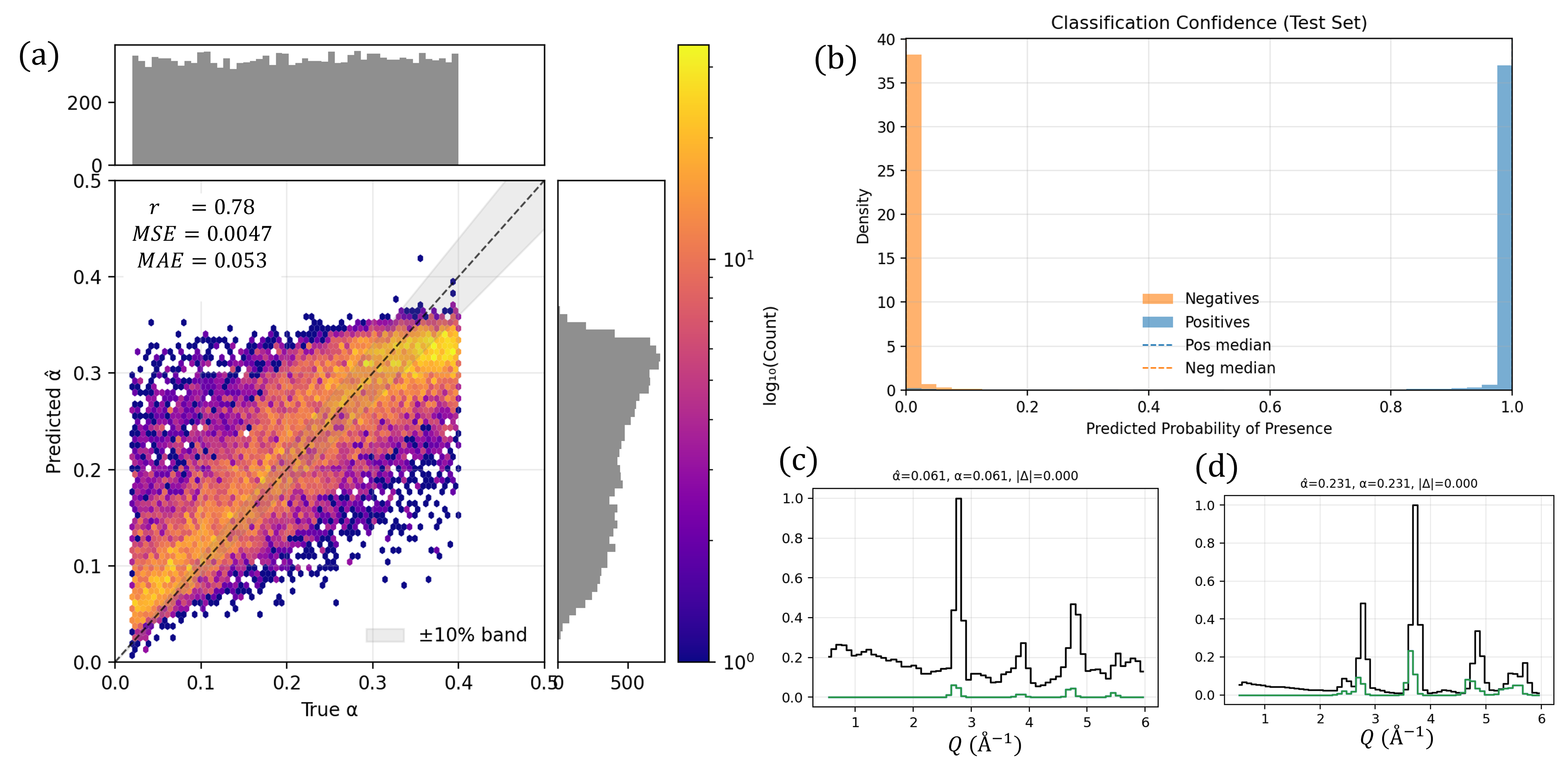}
\caption{Residual-histogram ranking model evaluated on held-out synthetic data. (a) Parity plot comparing predicted and true scale coefficients for candidates present in synthetic mixtures. (b) Predicted probability distributions for positive and negative examples. (c,d) Representative residual fingerprints (black) and candidate fingerprints scaled by the predicted coefficient (green).}
\label{fig:ml_performance}
\end{figure*}

We use a compact neural network that operates on paired coarse histograms---the residual fingerprint and a candidate fingerprint---together with a binary mask indicating the overlapping $Q$ range. The architecture combines lightweight one-dimensional convolutional feature extraction with a multi-head self-attention block, followed by pooling and fully connected layers. It outputs (i) a probability that the candidate phase is present and (ii) an approximate scale coefficient representing its contribution to the residual. The model is trained on synthetic mixtures generated with the \texttt{hfirestimate} neutron simulation framework \cite{Garlea2010HB2A,HFIRESTIMATE2025}, using database-derived structures with randomized lattice perturbations and noise. Crucially, because the network trains on an aggressively binned $Q$ grid rather than raw $2\theta$ or TOF profiles, it learns broad residual–candidate compatibility instead of instrument-specific resolution functions. This makes the model inherently instrument-agnostic and eliminates the need for retraining across different beamlines or radiation modalities. Figure~\ref{fig:ml_performance} shows representative held-out performance, illustrating that the model learns residual compatibility at the level of broad scattering features rather than requiring exact peak-by-peak alignment.


\subsection{Lattice nudging and refinement-based verification}

Database entries often differ from experimental conditions (temperature, composition or strain), shifting Bragg positions enough to destabilize refinement even when the correct structural prototype is present. After ML screening reduces the candidate set (typically to 10--20 phases), we therefore apply a fast lattice-nudging step to select a symmetry-consistent lattice variant that best aligns the candidate with the residual. In brief, we explore symmetry-allowed lattice distortions by sampling a low-dimensional ``$Q$-signature'' of representative low-index reflections and selecting a small diverse set of lattice variants for surrogate scoring (see Methods). The best-scoring variant is written as a new CIF and used to initialize GSAS-II verification.

Verification proceeds in two stages. First, each nudged candidate is refined directly against the residual curve with only its scale factor and lattice parameters allowed to vary within bounds; candidates are re-ranked by correlation between the refined profile and the residual in the experimental coordinate. Second, the top candidates are combined with the refined main phase in a conservative joint multiphase refinement against the raw histogram. After convergence, the impurity set is pruned by retaining the phase with the largest refined weight fraction and removing the rest, followed by a final polish refinement. This separation between rapid search and physically grounded verification provides interpretable outputs while keeping computation compatible with on-beamline use.


\begin{figure*}
    \centering
    \includegraphics[width=0.95\textwidth]{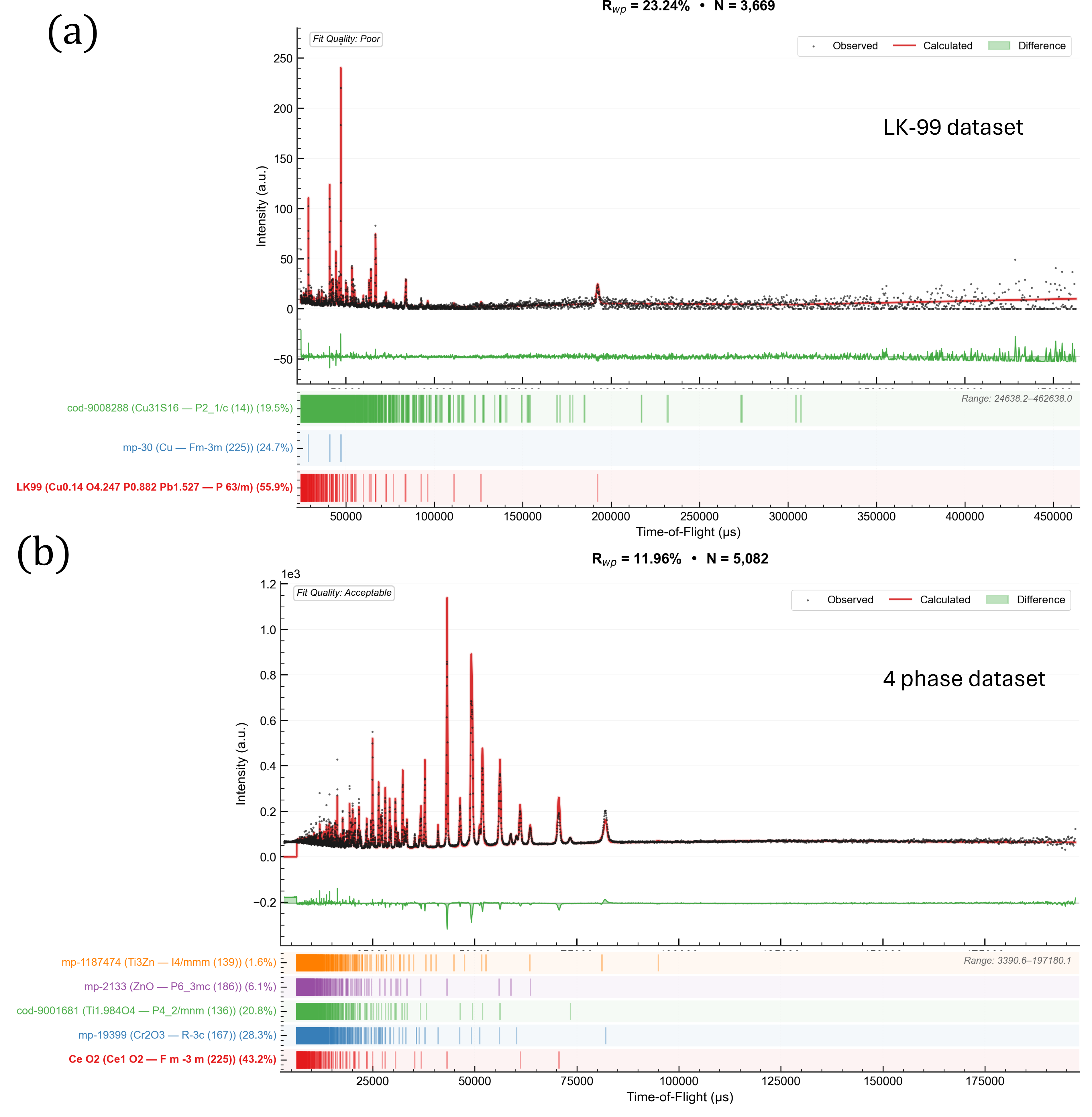}
    \caption{Representative GSAS-II fit plots produced by the pipeline. Black points: observed intensity; red: calculated profile; green: difference; tick marks: Bragg positions. (a) POWGEN LK-99: refined phases include LK-99, Cu and a \ce{Cu2S}-family phase. (b) POWGEN oxide mixture: refined phases include \ce{CeO2}, \ce{Cr2O3}, rutile-\ce{TiO2} and \ce{ZnO}, with an additional minor phase.}
    \label{fig:gsas_joint_fits}
\end{figure*}

\subsection{Benchmarking on synthetic mixtures}

We benchmarked the end-to-end workflow on a synthetic constant-wavelength neutron powder diffraction dataset of $N=18{,}491$ two-phase mixtures generated under beamline-realistic counting noise, peak overlap and modest lattice mismatch \cite{HFIRESTIMATE2025}. Using a family-level criterion that credits crystallographically equivalent database entries, the pipeline recovered the injected impurity phase in $15{,}513$ cases (83.9\%).

We also evaluated the composition-only mode in which no main-phase CIF is provided and the pipeline must first identify the dominant phase from an allowed element set. On this benchmark ($N=7{,}191$ mixtures), the correct main phase was recovered in 6{,}207 cases (86.3\%). Conditioned on correct main-phase recovery, the injected impurity phase was identified with a success rate of 5{,}558/6{,}207 (89.5\%). These end-to-end synthetic benchmark results are summarized in Table~\ref{tab:benchmark_summary}. These results evaluate the complete pipeline: success requires that the injected impurity survives screening, remains competitive after lattice nudging and is ultimately selected by the GSAS-II refinement and pruning procedure.

\subsection{Experimental case studies}
\label{sec:exp_case_studies}
We evaluated the workflow on experimental neutron powder diffraction datasets from ORNL instruments HB-2A (CW) and POWGEN (TOF) \cite{Calder2018ORNL_PowderSuite,Garlea2010HB2A,Huq2019POWGEN}. These cases include structured backgrounds, sample-environment scattering, peak overlap and imperfect starting models. Unless noted, runs use the standard beamline mode (measured histogram + instrument parameter file + main-phase CIF).

\paragraph{HB-2A (CW): container-phase identification under texture.}
Our first dataset is a CW diffraction pattern of \ce{Tb2Be2GeO7} measured on HB-2A, where the pattern contains additional reflections from the aluminum sample can. The container signal exhibits preferred orientation (texture), so relative reflection intensities deviate from ideal powder statistics. This case tests whether the workflow can (i) refine the main phase and background stably, (ii) isolate unexplained intensity into a residual, and (iii) identify a realistic sample-environment contaminant even when its intensity ratios are distorted by texture.

We provided the measured histogram, the GSAS-II instrument parameter file, the main-phase CIF for \ce{Tb2Be2GeO7}, and element constraints corresponding to the sample composition, with \ce{Al} included as a permitted sample-environment contaminant. After baseline refinement, the residual-based screening and verification stages ranked aluminum among the top impurity hypotheses and selected it in downstream refinement. The final model identifies fcc Al (space group $Fm\bar{3}m$, No.~225) despite texture-induced intensity distortions. 

\paragraph{POWGEN (TOF): LK-99 with strong impurities and lattice mismatch.}
Our second benchmark is a TOF POWGEN dataset collected on an LK-99 sample\cite{Zhang2024LK99PRMaterials}. This is a challenging case for automated impurity identification because the chemistry space is large, the measured intensity contains substantial impurity contributions, and many relevant phases have low symmetry with numerous weak and overlapping reflections. In such settings, baseline main-phase refinement can be imperfect: when a large fraction of peaks are not explained by the starting model, background and scale can trade off against unexplained intensity, yielding over- or under-subtracted residuals.

We provided the measured histogram, instrument parameter file, and the LK-99 main-phase CIF. The workflow recovered the two major impurity phases consistent with reported results\cite{Zhang2024LK99PRMaterials}: elemental Cu (fcc, $Fm\bar{3}m$, No.~225) and a \ce{Cu2S}-family phase (monoclinic, space group No.~14). In our final refined model, the dominant phases were LK-99 ($\sim$55.9~wt\%), Cu ($\sim$24.7~wt\%), and \ce{Cu31S16} ($\sim$19.5~wt\%; space group No.~14, in the same family as \ce{Cu2S}). For reference, the published refinement reported approximately 41.2\% (LK-99), 24.1\% (Cu), and 34.8\% (\ce{Cu2S})\cite{Zhang2024LK99PRMaterials}. While absolute fractions can differ due to instrument/modeling choices and database parameterization, the workflow automatically recovers the correct dominant impurity identities in a heavily multiphase TOF setting.

This dataset also highlights the importance of lattice nudging. The Cu entry in the Materials Project database (mp-30) has $a=\SI{3.577}{\angstrom}$, whereas the refined Cu lattice parameter in this sample is $a\approx \SI{3.616}{\angstrom}$ ( $a\sim$1\% difference). In direct GSAS-II refinement starting from the unmodified database CIF, this mismatch can be sufficient to prevent stable convergence for the Cu phase (often driving its refined scale factor toward zero), creating a realistic failure mode in manual screening. By explicitly accounting for lattice mismatch before refinement, the nudging stage mitigates this sensitivity and improves the reliability of downstream GSAS-II verification. The workflow additionally proposed a minor extra candidate phase not reported in the reference analysis; because the pipeline outputs an initialized GSAS-II project, such hypotheses can be rapidly audited and rejected based on peak-level evidence.

\paragraph{POWGEN (TOF): four-phase oxide mixture.}
Our third benchmark is a POWGEN TOF dataset containing a four-phase oxide mixture with known constituents \ce{CeO2}, \ce{TiO2}, \ce{Cr2O3}, and \ce{ZnO}. This example probes the pipeline’s ability to stabilize refinement and recover multiple phases in a realistic TOF profile with overlapping peaks and structured background, spanning both dominant ($\sim$40~wt\%) and minor ($\sim$few wt\%) components. Using the standard beamline mode with \ce{CeO2} supplied as the main phase, the final refined model retained all four expected phases, with refined weight fractions of approximately 43.2~wt\% (\ce{CeO2}, $Fm\bar{3}m$, No.~225), 20.8~wt\% (\ce{TiO2}, $P4_2/mnm$, No.~136), 28.3~wt\% (\ce{Cr2O3}, $R\bar{3}c$, No.~167), and 6.1~wt\% (\ce{ZnO}, $P6_3mc$, No.~186), plus a minor additional phase at the $\sim$1--2~wt\% level. The corresponding ground-truth mixture fractions for this sample were approximately 39\% (\ce{CeO2}), 25\% (\ce{TiO2}), 28\% (\ce{Cr2O3}), and 7.4\% (\ce{ZnO}). We emphasize that the purpose of this benchmark is to demonstrate reliable recovery of the correct phase set and a physically plausible multiphase refinement that users can directly audit and extend.

\paragraph{ Composition-only mode (no main-phase CIF).}
Although the primary beamline use case assumes a main-phase CIF is available, the workflow also supports an exploratory composition-only mode in which only an allowed element set is provided. On the above POWGEN dataset, this composition-only initialization recovered the correct dominant oxide phases and produced a stable multiphase refinement without requiring a user-supplied starting model. The accepted model contained \ce{CeO2} (mp-20194; $Fm\bar{3}m$, No.~225), \ce{Cr2O3} (mp-19399; $R\bar{3}c$, No.~167), and a \ce{TiO2} entry (cod-9001681; $P4_2/mnm$, No.~136), with refined weight fractions of approximately 44.6~wt\%, 30.4~wt\%, and 22.8~wt\%, respectively, along with a minor \ce{ZnCrO4} phase (mp-755896; $\sim$2.3~wt\%). It however missed the \ce{ZnO} phase.  This result illustrates that the workflow can bootstrap a physically plausible phase set directly from elemental constraints, which is useful for unlabeled datasets or exploratory synthesis where only the precursor element set is known.

\paragraph{PXRD experimental dataset}

In neutron diffraction, a common use case involves a known primary phase with an unknown impurity phase, since experiments at beamlines are typically performed after substantial preliminary characterization. In contrast, for benchtop powder X-ray diffraction (PXRD), even assigning a reliable reference phase can be nontrivial. 
Here, we benchmark phase-identification performance on experimental PXRD data from the RRUFF database and compare RADAR-PD against DARA \cite{Fei2026Dara}, which performs an exhaustive tree search over all candidate phases. Because RRUFF powder entries do not provide a direct, unambiguous mapping to a unique structure-database identifier, we first constructed a trusted reference subset by matching each DIF entry to crystallographic database candidates using composition and lattice compatibility. We then selected plausible labels via LLM-assisted diffraction comparison and retained only cases whose selected label passed Rietveld refinement quality criteria (Methods). This process yielded a trusted experimental subset of $n=291$ samples.

Predictions were scored using LLM-based phase equivalence against these trusted reference labels, so that slight off-stoichiometry, light-atom doping, or differences in cell convention do not count as failures. On this benchmark, RADAR-PD recovered the trusted reference phase in 232/291 samples (79.7\%), compared with 187/291 (64.3\%) for DARA (Table~\ref{tab:benchmark_summary}).

\newcolumntype{C}[1]{>{\centering\arraybackslash}p{#1}}

\begin{table*}[t]
\centering
\small
\caption{
Benchmark summary on synthetic neutron powder diffraction (NPD) mixtures and the experimental RRUFF PXRD dataset.
Synthetic mixtures evaluate impurity recovery in two-phase systems and end-to-end phase identification when only an allowed element set is provided.
For the experimental RRUFF PXRD benchmark, a trusted subset was constructed by catalog matching, LLM-assisted label selection, and Rietveld validation (Methods), yielding $N=291$ samples.
Predictions were counted as successful only if the proposed phase was verified by the LLM against the trusted reference label; missing DARA predictions were counted as failures.
For the experimental benchmark, runtime is reported as mean / median / P95 in minutes.
}
\label{tab:benchmark_summary}

\setlength{\tabcolsep}{0pt}
\renewcommand{\arraystretch}{1.15}

\newlength{\synwidth}
\newlength{\expwidth}
\setlength{\synwidth}{0.98\textwidth}
\setlength{\expwidth}{0.90\textwidth}

\newlength{\synA} 
\newlength{\synB} 
\newlength{\synC} 
\newlength{\synD} 
\setlength{\synA}{2.5cm}
\setlength{\synB}{6.9cm}
\setlength{\synC}{1.4cm}
\setlength{\synD}{3.4cm}

\newlength{\expA} 
\newlength{\expB} 
\newlength{\expC} 
\newlength{\expD} 
\setlength{\expA}{2.6cm}
\setlength{\expB}{1.2cm}
\setlength{\expC}{3.5cm}
\setlength{\expD}{4.8cm}

\begin{tabular*}{\synwidth}{@{\extracolsep{\fill}} C{\synA} C{\synB} C{\synC} C{\synD} @{}}
\toprule
\multicolumn{4}{@{}l}{\textit{Synthetic NPD two-phase mixtures}} \\
\midrule
\textbf{Setting} & \textbf{Task} & \textbf{$N$} & \textbf{Success} \\
\midrule
Two-phase  & Impurity recovery (dominant known) & 18,491 & 83.9\% (15,513/18,491) \\
Comp.-only & Dominant phase identification      & 7,191  & 86.3\% (6,207/7,191) \\
Comp.-only & Impurity recovery (given dominant) & 6,207  & 89.5\% (5,558/6,207) \\
\bottomrule
\end{tabular*}

\vspace{0.7em}

\begin{tabular*}{\expwidth}{@{\extracolsep{\fill}} C{\expA} C{\expB} C{\expC} C{\expD} @{}}
\toprule
\multicolumn{4}{@{}l}{\textit{Experimental RRUFF PXRD (trusted subset; phase-match task)}} \\
\midrule
\textbf{Method} & \textbf{$N$} & \textbf{Success} & \shortstack[c]{\textbf{Runtime} \\ \textbf{(mean / median / P95, min)}} \\
\midrule
DARA     & 291 & 64.3\% (187/291) & 84.8 / 16.0 / 427.6 \\
RADAR-PD & 291 & 79.7\% (232/291) & 19.0 / 9.9 / 58.2 \\
\bottomrule
\end{tabular*}

\end{table*}
\subsection{Performance and computational timing}

We next compared end-to-end runtime on the same 291-sample experimental PXRD benchmark used for accuracy evaluation. RADAR-PD required a mean of 19.0~min per sample and a median of 9.9~min, with a P95 runtime of 58.2~min and a maximum runtime of 535.6~min. DARA required a mean of 84.8~min per sample (4$\times$ higher than RADAR-PD) and a median of 16.0~min, with a substantially heavier tail (P95 427.6~min, $\sim$7.3$\times$ higher than RADAR-PD; maximum 1128.3~min). Thus, RADAR-PD was not only more accurate on this benchmark but also substantially faster. This improvement arises because RADAR-PD performs costly pattern calculations for only a small subset of selected candidates (typically 10--20), whereas DARA evaluates all candidates.

For RADAR-PD, GSAS-II refinement remains the dominant computational cost, while histogram screening and lattice nudging are comparatively lightweight and primarily serve to restrict refinement to a small shortlist of candidates. In practice, this enables rapid candidate identification, followed by refinement-based verification and quantification when additional computation time is available.
The implementation supports multiprocessing, allowing parallel GSAS-II verification across available CPU cores to reduce wall-clock time. For rapid on-shift triage, intermediate results are reported before refinement completes: a top-$k$ shortlist (default $k=5$) is displayed within the first minute using a lightweight multi-layer perceptron (MLP) re-ranker applied to candidate-level screening metrics. Lattice nudging and GSAS-II verification then proceed automatically to deliver quantitative phase fractions and an audit-ready GSAS-II project.

Overall, the workflow is optimized for near-immediate candidate identification to guide beamline decisions, followed by refinement-based quantification when additional computation time is available.
\subsection{Graphical user interface and interactive deployment}


\begin{figure*}[t]
    \centering
    \includegraphics[width=\textwidth]{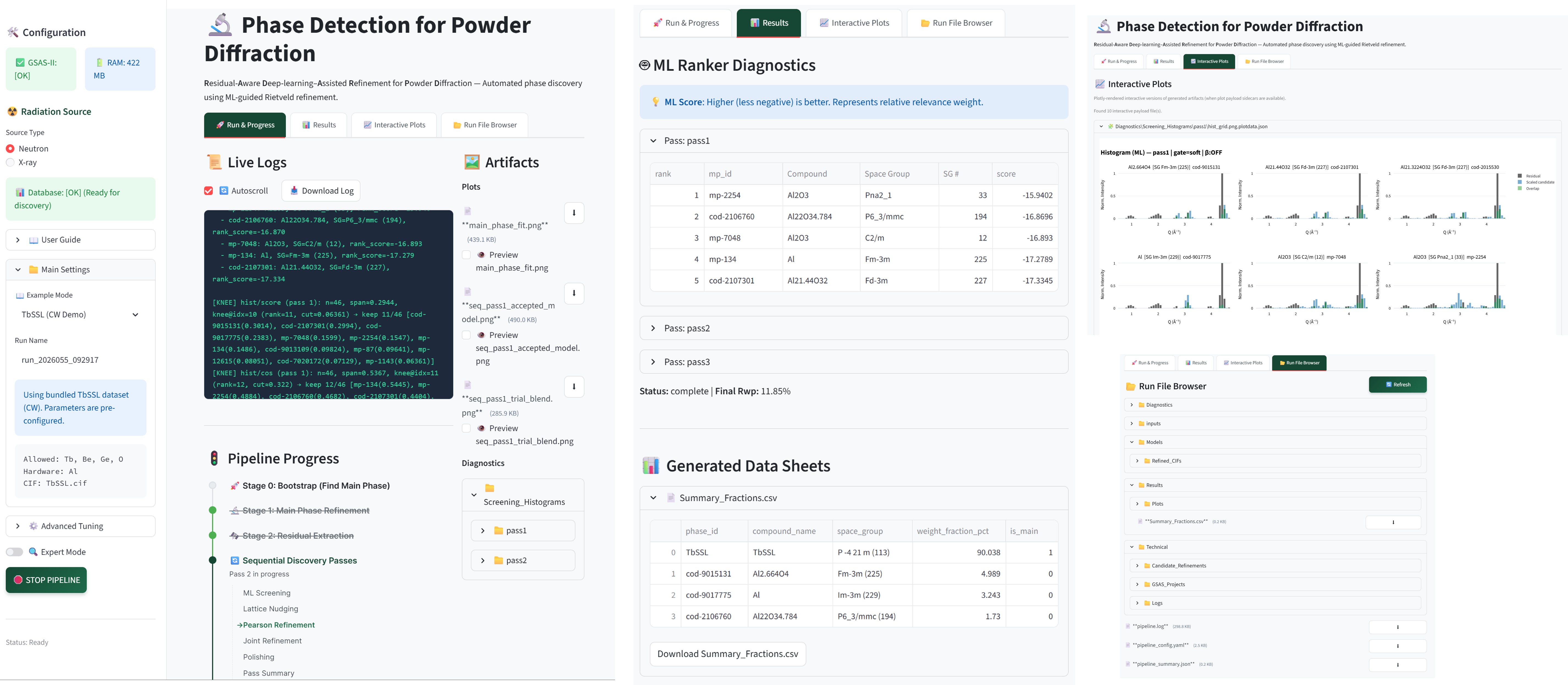}
    \caption{Graphical user interface (GUI) for RADAR-PD. Left: configuration and live pipeline progress during a run, including radiation-source selection, database readiness, logs, and sequential discovery stages. Center: results view showing ML ranker diagnostics and generated phase-fraction summaries. Right: interactive plots and run-file browser for inspecting candidate-screening histograms, refinement artifacts, and exported outputs. The interface supports real-time impurity triage, refinement auditing, and download of reproducible analysis artifacts across neutron and X-ray workflows.}
    \label{fig:gui}
\end{figure*}

To facilitate adoption beyond command-line workflows, we provide a graphical user interface (GUI) implementation of the pipeline (Fig.~\ref{fig:gui}). The GUI exposes the same residual-explanation workflow described above, including radiation modality selection (neutron or X-ray), instrument configuration (constant-wavelength or time-of-flight), elemental constraints, candidate screening parameters and refinement controls.

The interface provides real-time pipeline diagnostics, including stage-by-stage progress tracking, ML ranking summaries, lattice-nudging results and GSAS-II refinement metrics. Ranked impurity candidates are displayed together with space group information, database identifiers and ML relevance scores. Interactive plots allow users to inspect coarse $Q$ fingerprints, residual histograms and final joint refinement fits. Generated artifacts—including refined CIFs, candidate screening histograms, summary tables and GSAS-II project files—are organized for download to support reproducibility and downstream analysis.

Radiation modality is selected at runtime, with the software automatically switching between modality-specific reference catalogs and scattering factors. This enables the same interface to support neutron diffraction as well as laboratory and synchrotron X-ray beamlines without modifying the underlying workflow.

The GUI is currently hosted for public testing and is designed to be compatible with larger-scale deployment within facility computing environments. While the benchmarks presented here were executed programmatically, the GUI provides an interactive pathway for beamline users to perform impurity triage, audit candidate rankings and inspect refinement results without requiring direct scripting.


\section{Discussion}

Automated phase analysis remains a core barrier to autonomous materials workflows because it is an \emph{open-set} retrieval problem under strong distribution shift. Experimental patterns vary across instruments and modalities in resolution, background structure and noise, while reference structures come from evolving databases whose lattice parameters often differ from experimental conditions. These mismatches break exact-pattern search--match and destabilize refinement-based screening, forcing manual trial-and-error. RADAR-PD addresses this gap with a modality-aware propose--correct--verify workflow that produces refinement-grade, auditable outputs on beamline-realistic timescales.

The key design choice is to separate responsibilities between learning and physics. RADAR-PD uses a compact neural scorer operating on coarse $Q$-space fingerprints to rapidly rank candidates while remaining tolerant to modest peak shifts, multiphase overlap and resolution differences. This proposer stage reduces database-scale candidate sets to a tractable shortlist without relying on closed-set classification. Crucially, phase selection is not delegated to the network: candidates are accepted only after staged GSAS-II verification on the experimental historgam, yielding interpretable fit diagnostics, refined phase fractions and a complete refinement project for auditing.

Although motivated by neutron beamline workflows with a known dominant phase, RADAR-PD generalizes from \emph{residual explanation} to \emph{dominant-phase discovery}. With a baseline model, it performs a conservative refinement and iteratively explains residual intensity to recover additional phases; without one, the composition-only mode infers the dominant phase from elemental constraints and then enters the same residual-driven multiphase loop. Thus, impurity identification is not a special case but an operating point of a general phase-discovery and multiphase Rietveld-initialization engine.

A second contribution is lattice nudging, which mitigates a common refinement failure mode: database-to-experiment peak-position mismatch. Even with the correct prototype, modest lattice shifts (temperature, composition, strain) can suppress scales or derail convergence. Nudging searches a symmetry-consistent lattice neighborhood with a fast surrogate score to align candidates before GSAS-II verification, improving robustness—especially under peak overlap and imperfect subtraction. 

We validated RADAR-PD across modalities and problem settings. On large synthetic neutron benchmarks, the end-to-end pipeline recovers injected phases under realistic noise, overlap and lattice mismatch, demonstrating coordinated performance across screening, correction and refinement selection. Experimental CW and TOF neutron case studies show robustness to structured backgrounds, sample-environment scattering and imperfect starting models. Furthermore, on experimental RRUFF laboratory PXRD patterns, RADAR-PD demonstrates exceptional dominant-phase recovery using, vastly outperforming existing automated tools in accuracy and time.

Computationally, RADAR-PD is designed for rapid decisions without sacrificing auditability. The proposer stages (histogram screening and nudging) narrow hundreds to thousands of candidates to a shortlist in tens of seconds and enable early reporting of top candidates, while refinement-based verification---the dominant cost---can be parallelized and provides the physically grounded selection needed for quantitative use. Howver, performance is bounded by database coverage. Quantitative phase fractions can be biased when standard Rietveld assumptions are violated (e.g., preferred orientation or microabsorption); in such cases RADAR-PD is most reliable as a tool for rapid, inspectable identification and refinement initialization, with quantitative outputs interpreted alongside diagnostics.

Overall, RADAR-PD provides a practical interface between diffraction measurements and autonomous decision logic: fast, mismatch-tolerant hypothesis generation combined with physics-constrained verification, delivering reproducible, refinement-verified phase discovery across neutron and X-ray modalities.

\section{Methods}

\subsection{Inputs, operating modes and baseline refinement}

In the standard beamline mode, inputs comprise a measured diffraction histogram, a GSAS-II instrument parameter file, and a main-phase CIF. A user-provided element set constrains database retrieval for candidate phases; common sample-environment contaminants can be included by extending this set. To maximize stability across instruments and sample conditions, the baseline GSAS-II refinement is restricted to background coefficients, the histogram scale factor, and main-phase lattice parameters. The residual curve used for phase discovery is computed by subtracting the refined baseline model from the measured profile.

In an optional composition-only mode, no main-phase CIF is provided. In this setting, the total measured histogram is used to propose an initial dominant-phase hypothesis from the allowed element set, after which the same residual-based iterative discovery loop proceeds.

\subsection{Candidate retrieval and coarse $Q$ fingerprints}

Candidate structures are retrieved from locally stored crystallographic databases (Materials Project and the Crystallography Open Database)\cite{Grazulis2012COD,Jain2013MaterialsProject} under the user-provided element constraints. For fast, instrument-agnostic screening, both the experimental residual (or total histogram in composition-only initialization) and each candidate reference pattern are represented on a common momentum-transfer axis. Intensities are integrated into 64 fixed-width $Q$ bins spanning $0.5 < Q < 6~\text{\AA}^{-1}$. To accommodate experimental measurements with differing or narrower $Q$ limits, the measured signal is binned at this same resolution and padded with zeros to match the full 64-bin array. A binary overlap mask is then passed alongside the input to restrict the neural network's attention exclusively to the physically measured $Q$ range. This coarse representation is used only for ML screening and ranking; authoritative phase selection and quantification are performed by GSAS-II refinement.

\subsection{Stage 1 ML: residual-histogram compatibility model}

The first ML stage screens the element-constrained candidate set using a residual--candidate compatibility model operating on paired coarse $Q$ histograms (experimental residual and candidate reference), together with an overlap mask to handle differing $Q$ coverage between the experimental histogram and candidate pattern. The model outputs a candidate-level compatibility score and auxiliary agreement measures used to rank candidates for downstream verification. The model is intentionally compact to enable rapid CPU evaluation over hundreds of candidates. Detailed architectural information is provided in the Supplementary Material.

\subsection{Synthetic training data for the histogram model}

Training data for the residual-histogram model are generated synthetically using the \texttt{hfirestimate} neutron powder diffraction simulator\cite{Garlea2010HB2A,HFIRESTIMATE2025}. Two-phase mixtures are constructed by combining a dominant phase with a secondary phase at ratios $(1-x):x$, with $x$ spanning approximately 1\%--40\% impurity fractions. To emulate database--experiment mismatch, lattice parameters are perturbed by up to $\pm 3\%$. To emulate baseline-refinement error, the dominant phase is imperfectly subtracted prior to residual construction using a perturbed scale and additive noise. Each example consists of a residual fingerprint, a candidate fingerprint and an active-bin mask, with targets given by a binary presence label and a continuous scale coefficient $\alpha$ ($\alpha=x$ for present impurities and $\alpha=0$ for absent candidates).

\subsection{Stage 2 ML: MLP re-ranking and early top-$k$ reporting}
\label{sec:ml_screen_rerank}

To provide an early shortlist while the pipeline continues to lattice nudging and GSAS-II verification, we use a second ML stage: a lightweight MLP re-ranker. The re-ranker consumes the candidate-level metrics produced by the histogram screening stage and outputs a refined relevance score per candidate. Candidates are sorted by this score to generate the displayed top-$k$ shortlist (default $k=5$), which appears early in execution and does not require refinement.

\paragraph{Input features.}
Each candidate is represented by a fixed-length feature vector constructed from histogram-model outputs and simple list-context descriptors. Features include raw screening metrics (e.g., the histogram-model score and associated agreement/explained-intensity quantities), rank-aware features (candidate rank, normalized rank and ratios relative to the top candidate), and list-context features (within-list ranks/percentiles and local score gaps). This design enables the re-ranker to use both absolute evidence and relative structure within the screened candidate list.

\paragraph{Model.}
The re-ranker is a compact multilayer perceptron (MLP) with stacked fully connected layers, batch normalization, ReLU activations and dropout, producing a single scalar relevance score per candidate.

\subsection{Lattice nudging}

Lattice nudging aims to find a symmetry-consistent lattice variant for each shortlisted candidate that better matches peak positions under experimental conditions, without performing an expensive free search over all lattice parameters. For each candidate, crystal-system constraints are inferred from the space-group number to determine symmetry-allowed lattice degrees of freedom. Lattice-dependent peak motion is summarized using a 7-component ``$Q$-signature'' built from low-index reflections $(100),(010),(001),(110),(101),(011),(111)$ computed from the candidate base lattice. Around this base signature, target signatures are sampled within a bounded neighborhood (default $\pm 20\%$ per component) while enforcing small lattice distortions (default $\pm 3\%$ in lattice constants). A farthest-point strategy seeded at the base lattice selects a small diverse set of target signatures, which are then mapped back to symmetry-consistent lattice parameters.

Each representative lattice is scored against the experimental residual using a fast surrogate profile computed from a neutron/X-ray stick pattern (\texttt{pymatgen})\cite{Ong2013Pymatgen} rendered onto a uniform $Q$ grid with Gaussian broadening. Similarity is computed using cosine similarity in log-compressed intensity space,
\begin{equation}
\mathrm{score} =
\cos\!\left(
\log\!\left[1+\kappa\,\tilde{S}+\varepsilon\right],
\log\!\left[1+\kappa\,R+\varepsilon\right]
\right),
\end{equation}
where $\tilde{S}$ is the simulated profile normalized by its maximum, $R$ is the residual on the same grid, and $(\kappa,\varepsilon)$ control dynamic-range compression. The best-scoring lattice variant is written as a nudged CIF and used to initialize GSAS-II verification.

\subsection{GSAS-II verification, joint refinement and iterative phase selection}

For each nudged candidate, we perform a minimal residual-based GSAS-II refinement (candidate scale and lattice parameters only) to re-rank candidates in the experimental coordinate. The top candidates are then included in a joint refinement against the raw histogram together with the refined main phase, using a conservative staged schedule (background and scales, then lattice parameters). After convergence, we retain the candidate phase with the largest refined weight fraction and discard the remaining candidates, followed by a final polish refinement on the reduced model.

If multiple additional phases are requested, the baseline model is updated, the residual is recomputed, and the search--verify loop repeats until the requested count is reached or no further improvement is obtained under predefined thresholds.

\subsection{Software outputs}

The pipeline outputs (i) a GSAS-II project file containing the refined multiphase model, (ii) ranked candidate hypotheses with estimated phase fractions and fit statistics, and (iii) plots and logs for inspection and provenance tracking. Two operating modes are supported: a standard beamline mode using a user-provided main-phase model, and a composition-only mode that proposes a dominant phase from an allowed element set before entering the residual-based discovery loop.
\subsection{RRUFF PXRD benchmark and comparison to Dara}
\label{sec:methods_rruff_benchmark}

\paragraph{Trusted experimental reference-set construction.}
RRUFF powder entries do not provide a direct, unambiguous mapping to a unique structure-database identifier, so we first constructed a trusted experimental reference set rather than benchmarking exact database-ID recovery. We parsed 3,008 of 3,013 RRUFF DIF files and extracted unit-cell parameters, peak lists, and atom-block chemistry. Entries without a usable atom block were excluded from reference construction. We then matched each DIF entry against a combined candidate database assembled from the Crystallography Open Database and the Materials Project\cite{Grazulis2012COD,Jain2013MaterialsProject}, requiring exact agreement of the element set together with permutation-aware unit-cell agreement: sorted lattice lengths and sorted cell angles were each required to agree within 5\%. This yielded 1,520 RRUFF entries with at least one catalog candidate. Of these, 1,434 also had a paired experimental XY raw scan and were retained for downstream validation.

\paragraph{LLM-assisted label selection and refinement validation.}
For each retained entry, diffraction patterns were simulated from the matched catalog CIFs using \texttt{pymatgen}\cite{Ong2013Pymatgen}. We then used an Azure OpenAI GPT-4 model in a comparative selection setting, providing the experimental DIF metadata together with all matched catalog candidates and asking the model to choose the best-supported phase label (or none) based on peak-position agreement, relative intensities, lattice parameters, and phase-name consistency. Exact prompt text is given in the Supplementary Information. This procedure produced selected candidate labels for 1,430 of the 1,434 retained entries. The selected labels were then subjected to Rietveld refinement validation, and only samples with refinement quality $\mathrm{GOF}<1.5$ and $\mathrm{Rwp}<50$ were retained. This yielded the final trusted experimental benchmark subset of 291 RRUFF PXRD samples which were visually verified by looking at the fit plots.

\paragraph{Benchmark predictions and scoring.}
RADAR-PD and Dara were then run on the same 291 trusted samples using the same combined COD$+$Materials Project search space. All candidate phases returned by each method were evaluated rather than only the top-ranked prediction: for RADAR-PD, candidate phases were taken from the reported \texttt{top\_phases} list, and for Dara from the reported \texttt{top\_fraction} list. Each predicted candidate was compared against the trusted reference label set using a second Azure OpenAI GPT-4 prompt that judged powder-XRD phase equivalence from the trusted reference CIF and the predicted CIF. This verification prompt was designed to tolerate alternate settings, supercells, hydrogen/hydration differences, and symmetry-lowered or defect variants when they corresponded to the same powder-diffraction phase; the exact prompt and few-shot examples are provided in the Supplementary Information. A sample was counted as correct if any predicted candidate matched any trusted reference label. If Dara returned no candidate, that sample was counted as a failure.

\paragraph{Outputs and audit artifacts.}
This procedure yielded the final head-to-head benchmark reported in Table~\ref{tab:benchmark_summary}. In addition to CSV summaries, we generated per-sample audit PDFs showing the raw experimental PXRD pattern, the trusted reference simulation, the Dara prediction, the RADAR-PD prediction, and separate LLM reasoning pages for trusted-label selection and method-to-reference verification. Runtime statistics were computed from the per-sample wall-clock times emitted by each pipeline (Dara: \texttt{elapsed\_seconds}; RADAR-PD: \texttt{runtime\_seconds}).

\section*{Code availability}

The RADAR-PD implementation, including workflow orchestration, machine-learning screening models, lattice-nudging routines and GSAS-II verification scripts, is available at \url{https://github.com/LalitYadav07/Impurity_detection_GSAS_ver6}. 
An interactive deployment is available at \url{https://huggingface.co/spaces/Lalityadav07/phase_detection}.

\section*{Acknowledgment}

We thank Qiang Zhang and Cheng Li for providing the LK-99 and oxide testing data, and Sara Haravifard and Rabindranath Bag for providing the \ce{Tb2GeBe2O7} testing data. We also thank Marshall McDonnell, Yuanpeng Zhang, Dayton Kizzire, and Chen Zhang for useful discussions. Y.C. and M.D. were supported by the Scientific User Facilities Division, Office of Basic Energy Sciences, U.S. Department of Energy, under Contract No. DE-AC05-00OR22725 with UT-Battelle, LLC. 
A portion of this research used resources at the High Flux Isotope Reactor and the Spallation Neutron Source, DOE Office of Science User Facilities operated by Oak Ridge National Laboratory.
This material is based upon work supported by the U.S. Department of Energy, Office of Science, Office of Basic Energy Sciences, under Award No. ERKCS22 for the project ``ILLUMINE -- Intelligent Learning for Light Source and Neutron Source User Measurements Including Navigation and Experiment Steering.''

\bibliographystyle{unsrtnat}
\bibliography{Biblography}

@article{Rietveld1969,
  author  = {Rietveld, H. M.},
  title   = {A profile refinement method for nuclear and magnetic structures},
  journal = {Journal of Applied Crystallography},
  year    = {1969},
  volume  = {2},
  number  = {2},
  pages   = {65--71},
  doi     = {10.1107/S0021889869006558},
  url     = {https://doi.org/10.1107/S0021889869006558}
}

@article{McCusker1999RietveldGuidelines,
  author  = {McCusker, L. B. and Von Dreele, R. B. and Cox, D. E. and Lou{\"e}r, D. and Scardi, P.},
  title   = {Rietveld refinement guidelines},
  journal = {Journal of Applied Crystallography},
  year    = {1999},
  volume  = {32},
  number  = {1},
  pages   = {36--50},
  doi     = {10.1107/S0021889898009856},
  url     = {https://doi.org/10.1107/S0021889898009856}
}

@article{HillHoward1987NeutronQPA,
  author  = {Hill, R. J. and Howard, C. J.},
  title   = {Quantitative phase analysis from neutron powder diffraction data using the Rietveld method},
  journal = {Journal of Applied Crystallography},
  year    = {1987},
  volume  = {20},
  number  = {6},
  pages   = {467--474},
  doi     = {10.1107/S0021889887086199},
  url     = {https://doi.org/10.1107/S0021889887086199}
}

@article{TobyVonDreele2013GSASII,
  author  = {Toby, Brian H. and Von Dreele, Robert B.},
  title   = {{GSAS-II}: the genesis of a modern open-source all purpose crystallography software package},
  journal = {Journal of Applied Crystallography},
  year    = {2013},
  volume  = {46},
  number  = {2},
  pages   = {544--549},
  doi     = {10.1107/S0021889813003531},
  url     = {https://doi.org/10.1107/S0021889813003531}
}

@article{ODonnell2018GSASIIAPI,
  author  = {O'Donnell, Jackson H. and Von Dreele, Robert B. and Chan, Maria K. Y. and Toby, Brian H.},
  title   = {A scripting interface for {GSAS-II}},
  journal = {Journal of Applied Crystallography},
  year    = {2018},
  volume  = {51},
  number  = {4},
  pages   = {1244--1250},
  doi     = {10.1107/S1600576718008075},
  url     = {https://doi.org/10.1107/S1600576718008075}
}

@article{Lutterotti2019FPSM,
  author  = {Lutterotti, Luca and Pilli{\`e}re, Henry and Fontugne, Christophe and Boullay, Philippe and Chateigner, Daniel},
  title   = {Full-profile search--match by the Rietveld method},
  journal = {Journal of Applied Crystallography},
  year    = {2019},
  volume  = {52},
  number  = {3},
  pages   = {587--598},
  doi     = {10.1107/S160057671900342X},
  url     = {https://doi.org/10.1107/S160057671900342X}
}

@misc{CrystalImpactMatch,
  author       = {{Crystal Impact}},
  title        = {{Match!} -- Phase Analysis using Powder Diffraction},
  howpublished = {Computer software},
  url          = {https://www.crystalimpact.com/match/},
   year         = {2003},
}

@article{Degen2014HighScoreSuite,
  author  = {Degen, Thomas and Sadki, Mustapha and Bron, Egbert and K{\"o}nig, Uwe and N{\'e}nert, Gwilherm},
  title   = {The HighScore suite},
  journal = {Powder Diffraction},
  year    = {2014},
  volume  = {29},
  number  = {S2},
  pages   = {S13--S18},
  doi     = {10.1017/S0885715614000840},
  url     = {https://doi.org/10.1017/S0885715614000840}
}

@article{Kabekkodu2024PDF5plus,
  author  = {Kabekkodu, Soorya N. and Dosen, Anja and Blanton, Thomas N.},
  title   = {{PDF-5+}: a comprehensive Powder Diffraction File{\texttrademark} for materials characterization},
  journal = {Powder Diffraction},
  year    = {2024},
  volume  = {39},
  number  = {2},
  pages   = {47--59},
  doi     = {10.1017/S0885715624000150},
  url     = {https://doi.org/10.1017/S0885715624000150}
}

@article{Hellenbrandt2004ICSD,
  author  = {Hellenbrandt, Mariette},
  title   = {The Inorganic Crystal Structure Database ({ICSD})---Present and Future},
  journal = {Crystallography Reviews},
  year    = {2004},
  volume  = {10},
  number  = {1},
  pages   = {17--22},
  doi     = {10.1080/08893110410001664882},
  url     = {https://doi.org/10.1080/08893110410001664882}
}

@article{Zagorac2019ICSDDevelopments,
  author  = {Zagorac, D. and M{\"u}ller, H. and Ruehl, S. and Zagorac, J. and Rehme, S.},
  title   = {Recent developments in the Inorganic Crystal Structure Database: theoretical crystal structure data and related features},
  journal = {Journal of Applied Crystallography},
  year    = {2019},
  volume  = {52},
  number  = {5},
  pages   = {918--925},
  doi     = {10.1107/S160057671900997X},
  url     = {https://doi.org/10.1107/S160057671900997X}
}

@article{Grazulis2012COD,
  author  = {Gra{\v{z}}ulis, Saulius and Da{\v{s}}kevi{\v{c}}, Adriana and Merkys, Andrius and Chateigner, Daniel and Lutterotti, Luca and Quir{\'o}s, Miguel and Serebryanaya, Nadezhda R. and Moeck, Peter and Downs, Robert T. and Le Bail, Armel},
  title   = {Crystallography Open Database ({COD}): an open-access collection of crystal structures and platform for world-wide collaboration},
  journal = {Nucleic Acids Research},
  year    = {2012},
  volume  = {40},
  number  = {D1},
  pages   = {D420--D427},
  doi     = {10.1093/nar/gkr900},
  url     = {https://doi.org/10.1093/nar/gkr900}
}

@article{Jain2013MaterialsProject,
  author  = {Jain, Anubhav and Ong, Shyue Ping and Hautier, Geoffroy and Chen, Wei and Richards, William Davidson and Dacek, Stephen and Cholia, Shreyas and Gunter, Dan and Skinner, David and Ceder, Gerbrand and Persson, Kristin A.},
  title   = {Commentary: The Materials Project: A materials genome approach to accelerating materials innovation},
  journal = {APL Materials},
  year    = {2013},
  volume  = {1},
  number  = {1},
  pages   = {011002},
  doi     = {10.1063/1.4812323},
  url     = {https://doi.org/10.1063/1.4812323}
}

@article{BaptistaDeCastro2022XERUS,
  author  = {Baptista de Castro, Pedro and Terashima, Kensei and Esparza Echevarria, Miren Garbine and Takeya, Hiroyuki and Takano, Yoshihiko},
  title   = {{XERUS}: An Open-Source Tool for Quick {XRD} Phase Identification and Refinement Automation},
  journal = {Advanced Theory and Simulations},
  year    = {2022},
  volume  = {5},
  number  = {5},
  doi     = {10.1002/adts.202100588},
  url     = {https://doi.org/10.1002/adts.202100588}
}

@article{Fei2026Dara,
  author  = {Fei, Yuxing and McDermott, Matthew J. and Rom, Christopher L. and Wang, Shilong and Ceder, Gerbrand},
  title   = {Dara: Automated Multiple-Hypothesis Phase Identification and Refinement from Powder X-ray Diffraction},
  journal = {Chemistry of Materials},
  year    = {2026},
  volume  = {38},
  number  = {3},
  pages   = {1364--1376},
  doi     = {10.1021/acs.chemmater.5c02820},
  url     = {https://doi.org/10.1021/acs.chemmater.5c02820}
}

@article{Dong2021PQNet,
  author  = {Dong, Hongyang and Butler, Keith T. and Matras, Dorota and Price, Stephen W. T. and Odarchenko, Yaroslav and Khatry, Rahul and Thompson, Andrew and Middelkoop, Vesna and Jacques, Simon D. M. and Beale, Andrew M. and Vamvakeros, Antonis},
  title   = {A deep convolutional neural network for real-time full profile analysis of big powder diffraction data},
  journal = {npj Computational Materials},
  year    = {2021},
  volume  = {7},
  number  = {1},
  pages   = {74},
  doi     = {10.1038/s41524-021-00542-4},
  url     = {https://doi.org/10.1038/s41524-021-00542-4}
}

@article{Zhang2024CPICANN,
  author  = {Zhang, Shouyang and Cao, Bin and Su, Tianhao and Wu, Yue and Feng, Zhenjie and Xiong, Jie and Zhang, Tong-Yi},
  title   = {Crystallographic phase identifier of a convolutional self-attention neural network ({CPICANN}) on powder diffraction patterns},
  journal = {IUCrJ},
  year    = {2024},
  volume  = {11},
  number  = {4},
  pages   = {634--642},
  doi     = {10.1107/S2052252524005323},
  url     = {https://doi.org/10.1107/S2052252524005323}
}

@article{Lee2020SyntheticXRD_DLPhaseID,
  author  = {Lee, Jin-Woong and Park, Woon Bae and Lee, Jin Hee and Singh, Satendra Pal and Sohn, Kee-Sun},
  title   = {A deep-learning technique for phase identification in multiphase inorganic compounds using synthetic {XRD} powder patterns},
  journal = {Nature Communications},
  year    = {2020},
  volume  = {11},
  number  = {1},
  pages   = {86},
  doi     = {10.1038/s41467-019-13749-3},
  url     = {https://doi.org/10.1038/s41467-019-13749-3}
}

@article{Schuetzke2021IucrjTrainingRealism,
  author  = {Schuetzke, Jan and Benedix, Alexander and Mikut, Ralf and Reischl, Markus},
  title   = {Enhancing deep-learning training for phase identification in powder X-ray diffractograms},
  journal = {IUCrJ},
  year    = {2021},
  volume  = {8},
  number  = {3},
  pages   = {408--420},
  doi     = {10.1107/S2052252521002402},
  url     = {https://doi.org/10.1107/S2052252521002402}
}

@article{Cao2025XQueryer,
  author  = {Cao, Bin and Zheng, Zinan and Liu, Yang and Zhang, Longhan and Wong, Lawrence W-Y and Weng, Lu-Tao and Li, Jia and Li, Haoxiang and Zhang, Tong-Yi},
  title   = {{XQueryer}: an intelligent crystal structure identifier for powder X-ray diffraction},
  journal = {National Science Review},
  year    = {2025},
  volume  = {12},
  number  = {12},
  doi     = {10.1093/nsr/nwaf421},
  url     = {https://doi.org/10.1093/nsr/nwaf421}
}

@article{Allan2019BlueskyAhead,
  author  = {Allan, Daniel and Caswell, Thomas and Campbell, Stuart and Rakitin, Maksim},
  title   = {Bluesky's Ahead: A Multi-Facility Collaboration for an a la Carte Software Project for Data Acquisition and Management},
  journal = {Synchrotron Radiation News},
  year    = {2019},
  volume  = {32},
  number  = {3},
  pages   = {19--22},
  doi     = {10.1080/08940886.2019.1608121},
  url     = {https://doi.org/10.1080/08940886.2019.1608121}
}

@article{McDannald2022ANDiE,
  author  = {McDannald, Austin and Frontzek, Matthias and Savici, Andrei T. and Doucet, Mathieu and Rodriguez, Efrain E. and Meuse, Kate and Opsahl-Ong, Jessica and Samarov, Daniel and Takeuchi, Ichiro and Ratcliff, William and Kusne, A. Gilad},
  title   = {On-the-fly autonomous control of neutron diffraction via physics-informed Bayesian active learning},
  journal = {Applied Physics Reviews},
  year    = {2022},
  volume  = {9},
  number  = {2},
  pages   = {021408},
  doi     = {10.1063/5.0082956},
  url     = {https://doi.org/10.1063/5.0082956}
}

@article{Calder2018ORNL_PowderSuite,
  author  = {Calder, S. and An, K. and Boehler, R. and Dela Cruz, C. R. and Frontzek, M. D. and Guthrie, M. and Haberl, B. and Huq, A. and Kimber, S. A. J. and Liu, J. and Molaison, J. J. and Neuefeind, J. and Page, K. and dos Santos, A. M. and Taddei, K. M. and Tulk, C. and Tucker, M. G.},
  title   = {A suite-level review of the neutron powder diffraction instruments at Oak Ridge National Laboratory},
  journal = {Review of Scientific Instruments},
  year    = {2018},
  volume  = {89},
  number  = {9},
  pages   = {092701},
  doi     = {10.1063/1.5033906},
  url     = {https://doi.org/10.1063/1.5033906}
}

@article{Garlea2010HB2A,
  author  = {Garlea, V. O. and Chakoumakos, B. C. and Moore, S. A. and Taylor, G. B. and Chae, T. and Maples, R. G. and Riedel, R. A. and Lynn, G. W. and Selby, D. L.},
  title   = {The high-resolution powder diffractometer at the high flux isotope reactor},
  journal = {Applied Physics A},
  year    = {2010},
  volume  = {99},
  number  = {3},
  pages   = {531--535},
  doi     = {10.1007/s00339-010-5603-6},
  url     = {https://doi.org/10.1007/s00339-010-5603-6}
}

@article{Huq2019POWGEN,
  author  = {Huq, Ashfia and Kirkham, Melanie and Peterson, Peter F. and Hodges, Jason P. and Whitfield, Pamela S. and Page, Katharine and H{\"u}gle, Thomas and Iverson, Erik B. and Parizzi, Andre and Rennich, George},
  title   = {{POWGEN}: rebuild of a third-generation powder diffractometer at the Spallation Neutron Source},
  journal = {Journal of Applied Crystallography},
  year    = {2019},
  volume  = {52},
  number  = {5},
  pages   = {1189--1201},
  doi     = {10.1107/S160057671901121X},
  url     = {https://doi.org/10.1107/S160057671901121X}
}

@article{Ong2013Pymatgen,
  author  = {Ong, Shyue Ping and Richards, William Davidson and Jain, Anubhav and Hautier, Geoffroy and Kocher, Michael and Cholia, Shreyas and Gunter, Dan and Chevrier, Vincent L. and Persson, Kristin A. and Ceder, Gerbrand},
  title   = {Python Materials Genomics ({pymatgen}): A robust, open-source python library for materials analysis},
  journal = {Computational Materials Science},
  year    = {2013},
  volume  = {68},
  pages   = {314--319},
  doi     = {10.1016/j.commatsci.2012.10.028},
  url     = {https://doi.org/10.1016/j.commatsci.2012.10.028}
}

@article{HFIRESTIMATE2025,
  author  = {Paddison, Joseph A. M. and Calder, Stuart and Yahne, Danielle R. and Cochran, Malcolm J. and Chen, Si Athena and Frontzek, Matthias D. and Zhang, Yuanpeng},
  title   = {A planning tool for neutron powder diffraction experiments},
  journal = {Review of Scientific Instruments},
  year    = {2025},
  volume  = {96},
  number  = {11},
  pages   = {113901},
  doi     = {10.1063/5.0283549},
  url     = {https://doi.org/10.1063/5.0283549}
}

@article{Zhang2024LK99PRMaterials,
  author  = {Zhang, Qiang and Guan, Yingdong and Cheng, Yongqiang and Min, Lujin and Keum, Jong K. and Mao, Zhiqiang and Stone, Matthew B.},
  title   = {Structure and lattice excitations of the copper substituted lead oxyapatite {Pb9.06(7)Cu0.94(6)(PO3.92(4))6O0.96(3)}},
  journal = {Physical Review Materials},
  year    = {2024},
  volume  = {8},
  number  = {1},
  pages   = {014605},
  doi     = {10.1103/PhysRevMaterials.8.014605},
  url     = {https://doi.org/10.1103/PhysRevMaterials.8.014605}
}

@misc{ThayerILLUMINE_LCLStream,
  author       = {{Linac Coherent Light Source (LCLS), SLAC National Accelerator Laboratory}},
  title        = {{LCLStream}},
  howpublished = {Web page},
  year         = {2026},
  url          = {https://lcls.slac.stanford.edu/depts/data-systems/projects/lclstream},
  note         = {Accessed 2026-03-23}
}

@misc{LCLS_ILLUMINE,
  author       = {{Linac Coherent Light Source (LCLS), SLAC National Accelerator Laboratory}},
  title        = {{ILLUMINE}},
  howpublished = {Web page},
  year         = {2026},
  url          = {https://lcls.slac.stanford.edu/depts/data-systems/projects/illumine},
  note         = {Accessed 2026-03-23}
}

@misc{OLCF2024_StreamingPipeline,
  author       = {{Oak Ridge Leadership Computing Facility (OLCF)}},
  title        = {Standing up the Nation's Supercomputing Pipeline for Streaming Big Data in Real Time},
  howpublished = {Web page},
  year         = {2024},
  url          = {https://www.olcf.ornl.gov/2024/10/14/standing-up-the-nations-supercomputing-pipeline-for-streaming-big-data-in-real-time/},
  note         = {Accessed 2026-03-23}
}
\end{document}